# Robust and Hyper-Efficient Multi-dimensional Optical Fiber Semantic Communication

Yuxuan Xiong†, Ziwen Zhou†, Jixing Ren†, Jingze Liu, Zheng Gao, Ting Jiang, Xuchen Hua, Gengqi Yao, Yuqi Li, Mingming Zhang, Hao Wu*, Siqi Yan*, and Ming Tang*

**Abstract**

The growing demands of artificial intelligence and immersive media require communication beyond bit-level accuracy to meaning awareness. Conventional optical systems that focused on syntactic precision suffer significant inefficiencies. Here, we introduce a multi-dimensional semantic communication framework that bridges this gap by directly mapping high-level semantic features onto the orthogonal physical dimensions of light, frequency, polarization, and intensity, within a multimode fiber. This synergistic co-design of semantic logic and the photonic channel achieve an unprecedented equivalent spectral efficiency approaching 1000 bit/s/Hz. Moreover, it demonstrates profound resilience, maintaining high-fidelity reconstruction even when the physical-layer symbol error rate exceeds 36%, a condition under which conventional communication systems fail completely. Crucially, this deeply integrated co-design of semantic encoding and physical-layer modulation enables full semantic demodulation with only single-ended intensity detection, therefore significantly reducing system complexity and cost. This work establishes a validated pathway toward hyper-efficient, error-resilient optical networks for the next generation of data-intensive computing.

**Introduction**

The rapid proliferation of large-scale artificial intelligence, autonomous systems, and immersive media is driving an unprecedented surge in global information exchange [1]. Recent forecasts predict that worldwide data volume will exceed 500 zettabytes by 2029, with the vast majority comprising unstructured content like images and videos [2]. This explosion in unstructured data exposes a fundamental inefficiency in current networks, since over 90% of this information is often duplicated or semantically redundant [3,4]. Such inefficiency leads to severe penalties in bandwidth utilization and energy consumption, posing a significant challenge to the sustainability of global information infrastructure.

This challenge is particularly acute in high-throughput environments like data centers, which process over 70% of global data traffic [5]. Within these facilities, where transmission distances are typically under 1 km, communication systems engineered for bit-wise accuracy are fundamentally ill-equipped for the meaning-centric nature of modern data exchange [6]. This core mismatch between systems demanding bit-level accuracy and the semantically rich data they carry necessitates a paradigm shift [7,8]. Therefore, the growing demand for intelligent, high-capacity data exchange requires a new communication strategy that focuses on conveying meaningful information by extracting and transmitting high-level semantic features, rather than just raw bits [8-10].

Recent breakthroughs in deep learning are now catalyzing a fundamental shift away from bit-level transmission toward goal-oriented communication, a paradigm designed to convey semantic meaning rather than syntactic data [11]. By leveraging the powerful feature extraction and contextual understanding capabilities of deep neural networks, modern semantic architectures can distill raw data into its core meaning [7]. This approach enables the transmission of condensed semantic information rather than redundant data bits. The resulting paradigm offers dramatic reductions in data redundancy while yielding unprecedented improvements in both spectral and energy efficiency, particularly for bandwidth-intensive multimodal content such as text, images, and video.

While this communication paradigm has proven transformative in wireless domains [12-15], its potential in optical fiber, the bedrock of data infrastructure, has been largely unrealized. Optical channels offer immense bandwidth and stable, near-deterministic propagation, making them an ideal physical substrate for the reliable, high-dimensional mapping that semantic systems demand, particularly for intra–data center links. However, it has been a challenge to bridge the conceptual gap between high-level semantic features and the physical degrees of freedom of light. Conventional joint source-channel coding, for instance, fails to exploit the intrinsic alignment between data semantics and optical signal dimensions [16,17]. Furthermore, while preliminary attempts to embed semantics into multimode fibers (MMF) via frequency–text mappings are encouraging, they are constrained by a lack of deep semantic-driven encoding and an inability to harness the full multiplexing potential of MMF, limiting both expressive capability and robustness [18].

Here, we propose and experimentally validate a multi-dimensional multimode fiber semantic communication system (MMFSC) that resolves the fundamental challenge of integrating semantic intelligence with physical-layer photonics. The framework repurposes the MMF from a passive channel into an active physical-layer pre-processor. By harnessing the inherent modal dispersion and speckle characteristics of the fiber, our system performs an optical computation that projects the multi-dimensionally encoded semantic information onto a single, spatially-resolved intensity pattern. This deep integration of semantic encoding with the physical channel achieves an equivalent spectral efficiency approaching 1000 bit/s/Hz, representing a compression of several orders of magnitude over conventional protocols. Moreover, it maintains profound robustness, enabling high-fidelity visual reconstruction at a physical symbol error rate (SER) exceeding 36%. Furthermore, critically for practical adoption, this synergy radically simplifies the receiver to a single-ended intensity detector, fundamentally reducing system cost and complexity. The MMFSC architecture provides a solution for intelligent, resource-efficient optical networks that are intended to redefine performance in future data centers and edge-computing infrastructures.

## Results
### System Architecture and Working Principle

To address the escalating demands of bandwidth and latency in AI-driven infrastructures, we introduce MMFSC, a fundamentally new optical transmission architecture that encodes semantic content directly into the physical degrees of freedom of light. Its deep integration of semantic logic with the physical channel makes it ideal for the short-haul, high-volume traffic found in data centers, cloud computing platforms, and the immersive data streams of VR/AR. The framework is supported on a knowledge base (KB) shared by the transmitter and receiver, ensuring accurate coding and decoding of semantic information. The end-to-end process comprises a semantic encoder that compresses raw data into high-level features. A direct relationship is established between the meaning of a signal and the physical representation of light. These are then projected onto the frequency, polarization, and intensity of light. As this multi-dimensionally encoded signal propagates, the MMF not only serves as a low-loss transmission medium, but also functions as a deterministic physical decoder, where its inherent modal dispersion enables direct discrimination of multiplexed symbols, thus converting the complex, high-dimensional input field into a unique temporal intensity profile at the output. A single-ended intensity detector acts as the receiver, eliminating the need for complex coherent detection. Finally, the semantic decoder uses the KB to reconstruct the original content from this optically processed signature, ensuring high-fidelity recovery.

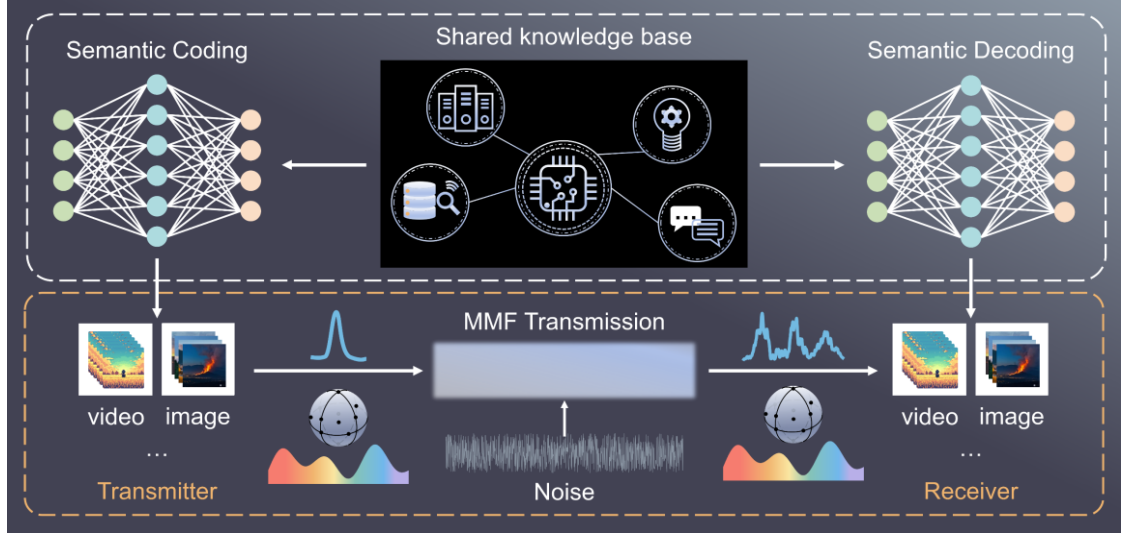

Fig. 1 **Conceptual framework of the MMFSC system.** The system encodes the semantic essence of data onto the physical dimensions of light including frequency, polarization, and intensity. The MMF is then leveraged as both the low-loss transmission medium and a physical-layer analog computer. Its intrinsic modal dispersion performs a pre-computation that projects the high-dimensional input state onto a unique temporal intensity fingerprint. This in-fiber processing is the key to receiver simplification, enabling the complete semantic state to be identified by a single, low-cost intensity detector and eliminating the need for complex coherent hardware. A semantic decoder then utilizes the optically computed fingerprint and a shared KB to reconstruct the original content with high fidelity for efficient intra–data center communication.

**Experimental setup**

The experimental setup to validate the MMFSC architecture was shown in Figure 2. Images and videos were semantically encoded through a shared knowledge base to form two 8-bit quantized low-rank feature matrices (see Methods). The experimental characterization of the modal dispersion of the 1 km MMF was performed to determine the mapping between this semantic information and the physical optical parameters. This characterization revealed a temporal broadening of approximately 8 ns, which dictates the system's temporal resolution and sets the operational parameters for avoiding inter-symbol interference.

The physical signal path begins with a narrow-linewidth (1 kHz) tunable laser. An I/Q modulator and an arbitrary waveform generator (AWG) modulate the frequency and intensity of the light according to the semantic matrices, while a polarization modulator (PM) sets the state of polarization (SOP). After amplification by an erbium-doped fiber amplifier (EDFA), the signal is launched into the 1-km MMF (105/125 μm) at a power below the nonlinear threshold. Based on the dispersion measurement, the pulse rate was set to 100 MHz with 1 ns and 0.75 ns pulse width. At the receiver, a 40 GHz photodetector and a high-speed digital storage oscilloscope (DSO) digitize the dispersed signal. The captured waveforms are then processed to decode the semantic information via the shared KB, enabling high-fidelity content reconstruction.

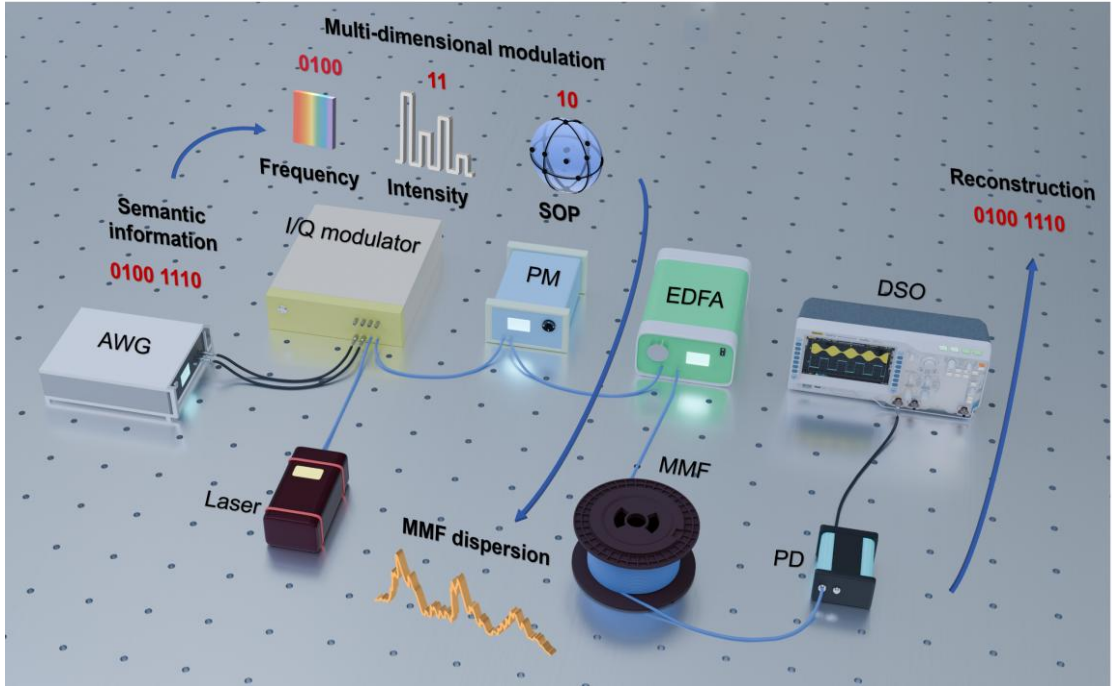

Fig. 2 **Experimental setup of the MMFSC system.** At the transmitter, semantic feature matrices generated from source data drive an AWG. The AWG controls an I/Q modulator and a PM, modulating the frequency, intensity, and SOP of light emitted from a narrow-linewidth tunable laser. An EDFA boosts the signal, which is then launched into a 1-km step-index MMF. At the receiver, a high-speed PD captures the dispersed optical signal, and a DSO digitizes the temporal waveform. Subsequent processing analyzes these waveforms to recover the encoded parameters, which are then semantically decoded using the shared KB to reconstruct the original content.

**Optical performance: computational resolution of the physical layer**

The information capacity of the MMFSC system is intrinsically governed by the computational resolution of the MMF. This resolution is a measure of the fiber's ability to map distinct multi-dimensional input symbols onto uniquely identifiable temporal output signatures. The physical mapping process relies on the MMF as a computing engine. An input optical symbol, characterized by its frequency, SOP, and intensity, excites a unique superposition of the $N$ guided modes with different effective refractive index $n_{eff,i}$. Their coherent interference generates a highly specific spatial speckle pattern. At the end facet, the intensity $I(x,y,L)$, at a distance $L$ from the input at position $(x,y)$, is described by [19]:

$$I(x,y,L) = \left| \sum_{i=1}^{N} a_i \hat{e}_i(x,y) e^{j\frac{2\pi}{\lambda}n_{eff,i}L} \right|^2 \quad (1)$$

Here, $\hat{e}_i(x,y)$ represents unit electric field. The complex excitation coefficients $a_i$ are a direct function of the input intensity and SOP, while the wavelength λ governs the relative phase between modes. Consequently, any change in the input symbol creates a deterministically different speckle pattern. The fiber's inherent intermodal dispersion then serializes this rich spatial information into a unique one-dimensional temporal waveform [20].

The deterministic physical transformation enables a highly simplified receiver where demodulation is implemented as a template-matching algorithm. A reference dictionary of these temporal "fingerprints" is recorded during calibration, and an incoming signal is identified by finding the entry with the maximum Pearson correlation. To quantify the limits of this in-fiber computation, the minimum separation required between states in each dimension was systematically characterized.

The spectral resolution was investigated by measuring the SER for a 256-symbol dictionary (equivalent to an 8-bit data structure). As shown in Figure 3a, the dispersion profiles for different frequencies are clearly distinguishable when the spacing exceeds 1 MHz, enabling error-free symbol discrimination. The spectral resolution limit, where the channel's properties begin to induce errors, was identified by progressively reducing this frequency interval. A critical trade-off emerges between spectral resolution and the optical pulse duration, as detailed in Figure 3b. Shorter pulses induce higher spectral crosstalk between adjacent symbols, leading to an increased SER for any given frequency spacing. This effect becomes particularly acute at a narrow 100 kHz channel separation, where the SER reaches 16% for a 1 ns pulse and escalates to 36% for a 0.75 ns pulse. The level of degradation would typically render a conventional multiplexed system inoperable. Consequently, these measurements establish an operational resolution threshold of 400 kHz for 1 ns pulses and 600 kHz for 0.75 ns pulses in a conventional frequency-encoded fiber communication system with 0-SER. And the results indicate that further compression of frequency interval is possible.

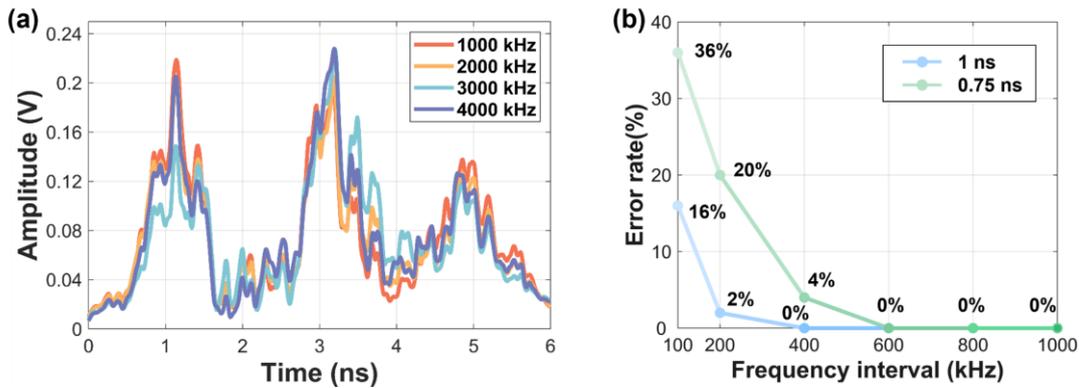

Fig. 3 **Spectral resolution limits and frequency multiplexing capacity of the MMFSC**.
**a,** Dispersion profiles for frequency-encoded symbols are fully resolved at a 1 MHz channel spacing around a 1550 nm central wavelength. **b,** the relationship between SER, pulse width, and frequency spacing quantifies the operational limits for frequency-domain multiplexing.

In the polarization domain, distinct dispersion curves for different SOPs were experimentally resolved, as shown in Figure 4, achieving an angular separation of $\pi/2048$ on the Poincare sphere. This high-resolution control enables the addressing of over 1.7 million distinct SOPs (See Method), a capacity that fundamentally expands the encoding state space beyond the two orthogonal states used in conventional polarization-division multiplexing.

Furthermore, modulation of the third orthogonal dimension, intensity, was implemented using a four-level pulse amplitude modulation (PAM-4) format, as illustrated in Figure 5. This approach produced four distinct and well-separated dispersion profiles, confirming intensity as an independent modulation axis capable of error-free operation.

Collectively, this systematic characterization quantifies the transfer function of the MMF when used for optical computation. The measured resolution limits in each dimension establish the fundamental capacity envelope of the MMFSC system. More critically, this detailed physical-layer analysis transforms the fiber channel into a deterministic operator whose computational properties are now fully defined. With known sensitivities to frequency, polarization, and intensity, this framework serves as the physical hardware specification for the semantic encoding logic. It forges a direct link between the fiber's intrinsic capacity for physical-state transformation and the achievable fidelity of the communicated meaning.

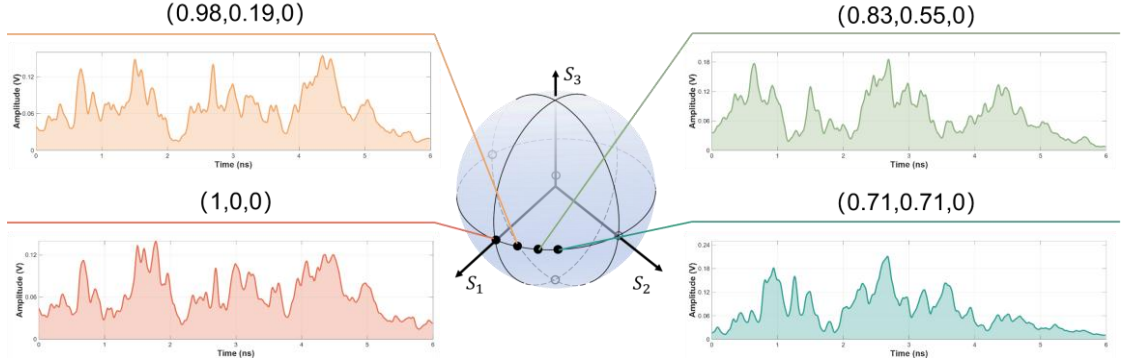

Fig. 4 **Resolution limits and capacity analysis of the MMFSC in polarization domains**. High-resolution polarization encoding demonstrated by distinct dispersion profiles for SOPs separated by small angular distances (0 to 3π/32) on the Poincare sphere from the reference (1,0,0).

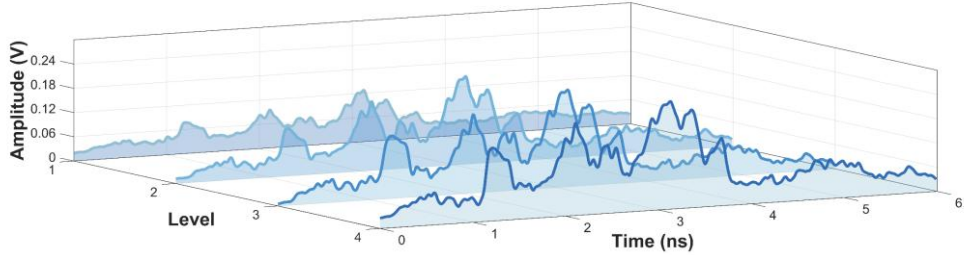

Fig. 5 **Resolution limits and capacity analysis of the MMFSC in intensity domains**. Four distinct intensity levels, implemented via PAM-4 generate clearly separable dispersion profiles, confirming intensity as a robust and independent modulation axis.

**Spectral Efficiency and Robust Transmission in Multidimensional Multiplexing.**

To quantify the information capacity of the MMFSC system, an analytical framework is introduced that unifies the gains from semantic compression with those from multidimensional optical encoding. Central to this framework is the definition of an equivalent spectral efficiency (ESE), a metric that captures the effective data throughput per unit of bandwidth. The ESE is designed to encapsulate the synergistic benefits arising from both the data reduction ratio at the semantic layer and the expansion of the symbol space enabled by multiplexing across frequency, polarization, and intensity. The ESE is defined in Eq. (1), where $R_s$ and $\Delta f$ are the transmission rate and frequency interval, respectively.

$$ESE = \frac{R_s \cdot \log_2 N}{\Delta f \cdot N_f} \quad (2)$$

The total dictionary size $N$ is determined by the product of the number of resolvable states in each optical dimension:

$$N = N_f \cdot N_{SOP} \cdot N_I \quad (3)$$

where $N_f$, $N_{SOP}$ and $N_I$ represent the number of distinct frequencies, SOPs, and intensity levels, respectively. Each unique combination of these optical parameters defines a symbol within the dictionary, enabling high-dimensional encoding.

This combinatorial structure allows the system to flexibly balance between robustness and spectral efficiency. With a symbol rate of 100 M and a frequency spacing of 100 kHz, distinct operational configurations demonstrate this adaptability. For instance, a balanced configuration with $N_f = N_{SOP} = 8, N_I = 4$ set achieving 1000-bit/s/Hz ESE with perfect error-free operation. In contrast, if the multiplexing of frequency and polarization are prioritized, a high-capacity setting with $N_f = N_{SOP} = 16, N_I = 1$ attains up to 500-bit/s/Hz while maintaining reliable performance with increased sensitivity to channel impairments.

The MMFSC system demonstrates a profound advantage in spectral efficiency. Operating at a symbol rate of 100 M, each image is semantically compressed to a compact 8808-byte representation, enabling a transmission rate of 11350 images per second. This entire data stream consumes a mere 16 kHz of spectral bandwidth, achieving an equivalent frame spectral efficiency of 0.71 frame/s/Hz. For a direct comparison, a conventional system transmitting the same number of uncompressed images would require a bandwidth on the order of several gigahertz. These results experimentally validate that the MMFSC architecture provides a hyper-efficient and scalable framework, uniquely capable of optimizing the trade-off between transmission rate and robustness to meet diverse application demands.

**Semantic communication with noise resilience and high-fidelity reconstruction**

The semantic-aware architecture of the MMFSC system offers a fundamentally different approach to error tolerance compared to conventional bit-level transmission. To evaluate this advantage, we employ both perceptual and objective metrics, LPIPS and PSNR, to quantify reconstruction quality under varying noise conditions. The LPIPS metric [21] closely mimics human visual perception by capturing both overall image quality and subtle detail loss perceptible to human observers. It exhibits an inverse relationship with perceived quality, meaning that lower scores indicate better perceptual fidelity. The PSNR, in contrast, provides an objective, pixel-level quality assessment through mean squared error computation between reference and reconstructed images. PSNR shows a direct positive correlation with image quality, where higher values reflect better-preserved image integrity and reduced distortion.

The semantic communication framework demonstrates profound resilience to channel-induced errors. Under a 1 ns pulse width condition that induces a 16% SER, the MMFSC system preserves high-fidelity image reconstruction, as shown in Figure 6. It yields favorable LPIPS scores of 0.2739,

0.3073, and 0.1715, while the PSNR experiences only a marginal decrease from an average of 27.1 dB to 25.6 dB. In stark contrast, a traditional communication (TC) system fails under these identical conditions. Its perceptual quality undergoes catastrophic degradation, reflected in LPIPS scores deteriorating to 0.6147, 0.5187, and 0.6337, and a PSNR that plummets from ~21.8 dB to an average of just 12.9 dB.

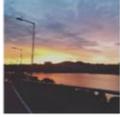

Fig. 6 **Robustness of semantic communication against channel-induced errors with 1 ns pulse width.** Comparative visualization of image reconstruction quality for the MMFSC framework versus a TC system under identical SER. Operating at a transmission rate of 100 M with a 1 ns input pulse width, the semantic approach preserves high perceptual fidelity where the conventional method exhibits catastrophic degradation. This visual evidence highlights the resilience endowed by prioritizing semantic content over bit-level accuracy.

The operational superiority of the semantic framework is starkly amplified under more stringent transmission conditions. Employing a 0.75 ns input pulse, which elevates the SER to 36%, the MMFSC system demonstrates profound resilience against this severe channel degradation. Its PSNR values only moderately decrease from an initial range of 21.53-27.74 dB to a final range of 18.37-25.08 dB, thereby preserving high-fidelity, perceptually coherent images, as shown in Figure 7. By contrast, the conventional bit-oriented framework suffers a catastrophic collapse under these identical conditions, rendering its transmitted images unrecognizable. This failure is quantified by LPIPS values escalating to 0.8839, 0.8321, and 0.9099, while the PSNR plummets from a baseline of 18.81 dB to an average of just 9.06 dB. Significantly, at the 36% SER that induces complete system failure for conventional methods, the semantic framework continues to reconstruct visually intelligible content. The profound immunity to channel noise signals a paradigm shift. By prioritizing semantic integrity over bit-level fidelity, system performance is fundamentally decoupled from traditional error-rate dependencies. This decoupling enables reliable information delivery in regimes previously considered inoperable. The intrinsic robustness is not limited to static data. The efficacy of the framework was further validated through the transmission of video sequences under a channel SER of 16%. Even at this significant error rate, the MMFSC system maintained high-fidelity video reconstruction with fluid motion and minimal perceptual distortion. The performance is quantified by a LPIPS score of 0.3285 and a PSNR of 26.50 dB. In contrast, under the same conditions, the TC system suffered a catastrophic loss of semantic integrity, with its LPIPS deteriorating to 0.7897 and its PSNR plummeting to 12.76 dB. As shown in Figure 8 and Supplementary Information, it demonstrates that the principle of semantic resilience extends robustly to dynamic, temporally correlated data streams.

![Fig. 7 comparison grid]

Fig. 7 **Semantic resilience under 0.75 ns pulse width condition.** Comparative visualization of MMFSC and traditional TC performance under high channel noise, induced by a 0.75 ns input pulse width at a 100 M transmission rate. At a 36% SER where the TC system experiences complete signal collapse, the MMFSC framework continues to reconstruct intelligible visual content. The result directly illustrates the profound noise immunity of the semantic approach, which maintains information integrity in conditions where conventional methods fail entirely.

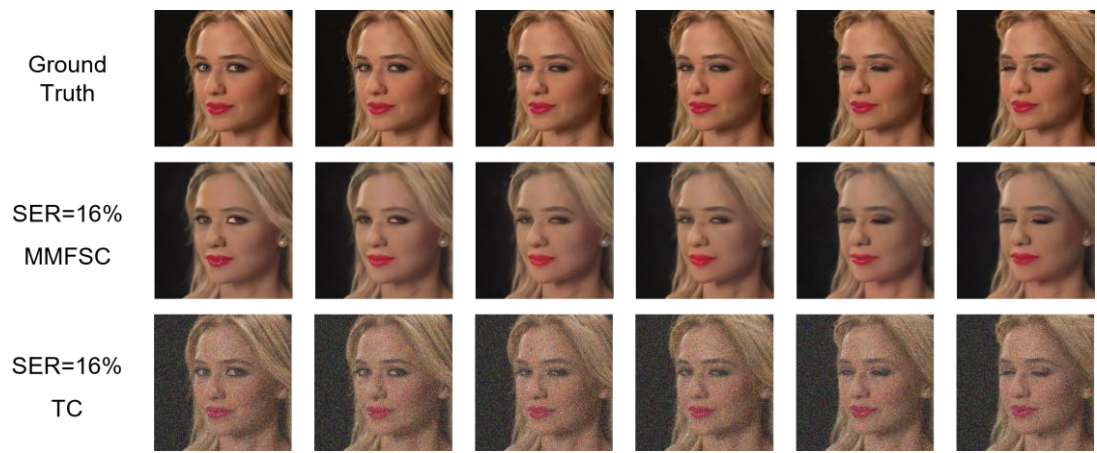

Fig. 8 **Superior fidelity of MMFSC for video transmission in a high-noise environment.** At a 16% SER, the MMFSC system successfully reconstructs a high-fidelity video frame, demonstrating its resilience to severe channel noise. In contrast, the frame transmitted via TC system is completely obscured by errors, highlighting the fundamental performance advantage of the semantic approach for dynamic information transfer.

## Discussion

In this work, we have demonstrated a multi-dimensional semantic communication system that fundamentally redefines the relationship between the physical transmission medium and the information it carries. The physical properties of the MMF are not a limitation to be overcome, but are actively leveraged as a computational element. Compressed semantic meaning are encoded onto the multi-dimensional state-space of light, including its frequency, polarization, and intensity. As it propagates, the MMF performs an analog pre-computation. The modal interference and intermodal dispersion project the high-dimensional input state onto a unique one-dimensional temporal intensity signature. It allows the complete semantic state to be recovered using only a single-ended intensity detector, drastically lowering the complexity and cost of the optical front-end.

The semantic-driven architecture, enabled by in-fiber computation, fundamentally decouples performance from traditional bit-error-rate dependencies. It prioritizes the preservation of meaning, allowing it to maintain intelligible communication in high-noise regimes where conventional methods collapse. The visual data is successfully reconstructed even when the physical-layer SER exceeds 36%. Furthermore, by transmitting only the essential semantic information, the system achieves an equivalent spectral efficiency approaching 1000-bit/s/Hz. The results establish a proven pathway for a new smart, resource-efficient optical interconnect technology that promises to redefine performance in data-intensive environments.

Crucially, this paradigm-shifting performance stems from a re-architecture of the communication task, not from exotic hardware. The in-fiber computation directly leads to the simplicity of hardware, which offloads the complex demodulation task from the electronic receiver to the physical channel. Therefore, the system is architecturally aligned with existing high-speed optoelectronics, using standard tunable lasers, polarization controllers [22,23], and modulators [24]. This compatibility ensures the deployment within current transceiver designs. Furthermore, the framework is inherently scalable. Future advances in optical device speed will directly enhance the computational resolution of the fiber, indicating potential for terabit-per-second semantic communication [25,26].

Beyond immediate applications, the principle of embedding computation within the physical layer opens new research directions in edge computing [27-30], neuromorphic photonics, and autonomous vehicle networks [31,32]. This work provides foundational insights into physics-aware system design and creates new opportunities in semantic-channel coding theory. Ultimately, it establishes a new design principle: by leveraging the transmission medium itself as a co-processor, the established boundaries of communication efficiency and robustness can be fundamentally broken.

## Methods
### Semantic encoding

Conventional communication systems, designed to achieve bit-level fidelity, treat all transmitted data with uniform importance. This approach is intrinsically inefficient for visual data, where considerable pixel-level variations can exist without altering the essential semantic content. Such a paradigm not only leads to excessive bandwidth consumption but also exhibits a critical vulnerability: a single bit-flip can introduce significant artifacts or lead to systemic failure.

To overcome these fundamental limitations, our optical semantic communication framework introduces a distinct encoding methodology, the architecture of which is contrasted with conventional systems in Figure 9. The focal point of the proposed methodology is the conversion of visual data into a sparse, low-rank matrix representation through image inversion, a process that efficiently captures the core visual information. For video sequences, we have developed an interframe compression strategy that preserves semantic continuity while substantially mitigating temporal data redundancy. This method moves beyond mere pixel-by-pixel replication, encoding the evolving meaning of the visual scene to ensure robust and efficient optical transmission.

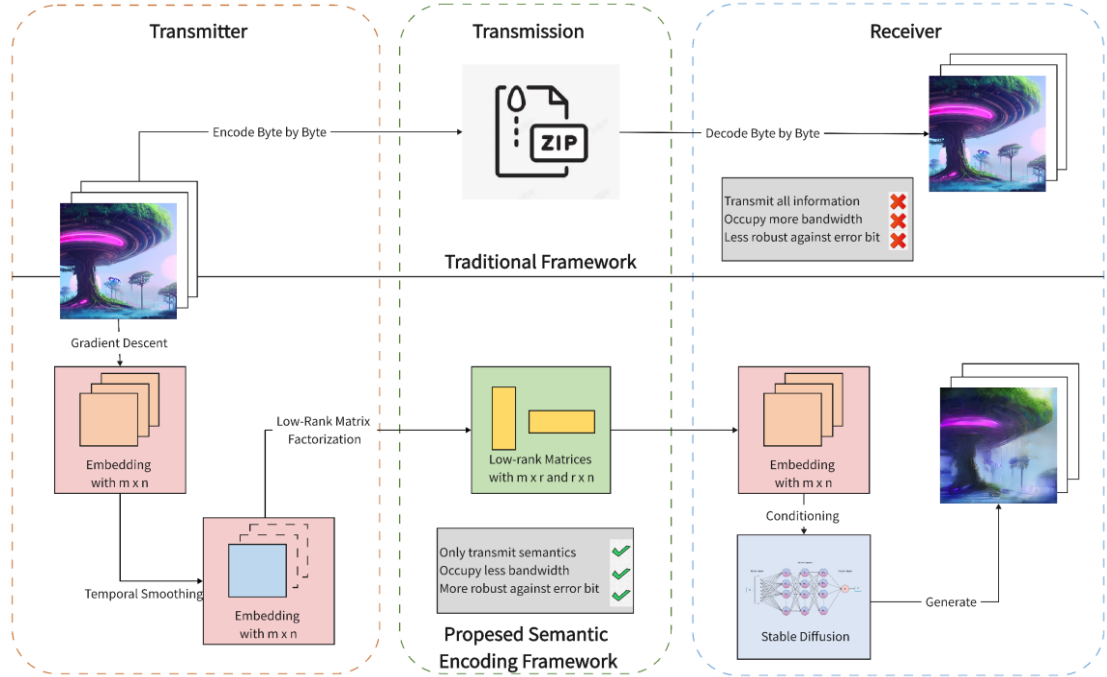

Fig. 9 **Semantic encoding architecture.** The system converts visual inputs into compressed semantic representations via text prompt inversion and low-rank factorization, enabling efficient transmission while maintaining reconstruction quality through generative decoding.

To fundamentally reduce transmission cost, the proposed framework encodes images not as arrays of pixels, but as compact, low-rank matrices derived from their semantic essence. This is achieved by leveraging a text-to-image generative model, Stable Diffusion, where the conditioning prompt required for image synthesis effectively serves as the semantic representation. While traditional image formats demand hundreds of kilobytes, this semantic proxy, captured in its low-rank form, typically requires only a few kilobytes, thereby realizing a significant compression ratio while preserving high-level visual meaning.

The generation of this semantic representation is accomplished via a two-stage image inversion protocol [33]. Initially, a $512 \times 512 \times 3$ image is inverted into a $77 \times 1024$ embedding through a gradient-descent optimization. This dense embedding is then compressed into two low-rank matrices of size $77 \times 8$ and $8 \times 1024$, suitable for efficient optical transmission. Recognizing that discrete text descriptions cannot ensure pixel-level fidelity, the optimization operates within a continuous embedding space. An embedding is iteratively refined by minimizing a loss function between the target image and the image generated by the model. To render this optimization tractable, the process employs SD-Turbo, a single-step denoising variant of Stable Diffusion that simplifies gradient computation, and directly optimizes the prompt embedding to circumvent the non-differentiability of the CLIP text encoder. Convergence is further accelerated by initializing the diffusion process not with random noise, but with a noised latent vector from a self-encoding of the target image (or the preceding frame in video sequences), which substantially reduces the latent distance the optimization must traverse. The final objective function is a composite of mean squared error (MSE) and the learned perceptual image patch similarity (LPIPS) metric, ensuring both pixel-level accuracy and perceptual realism in the reconstructed visual data [20].

$$\mathcal{D} = \alpha \cdot \mathcal{D}_{MSE} + (1-\alpha) \cdot \mathcal{D}_{LPIPS} \qquad (4)$$

where $\mathcal{D}_{MSE}$ is the reconstruction loss, $\mathcal{D}_{LPIPS}$ is the perceptual loss, and $\alpha$ balances fidelity and perceptual quality. In our experiment, $\alpha$ is set to 0.8. LPIPS is a perceptual metric designed to evaluate the visual similarity between two images in a way that aligns closely with human perception. Formally, given two images $x$ and $x'$, $\mathcal{D}_{LPIPS}$ is defined as:

$$\mathcal{D}_{LPIPS}(x, x') = \sum_{l} w_l \cdot \|\phi_l(x) - \phi_l(x')\|_2^2 \qquad (5)$$

where $\phi_l(x)$ represents the feature activation at the $l-th$ layer of a pre-trained deep neural network VGG and $w_l$ are learned weights that reflect the perceptual sensitivity of each layer.

To control the bitrate, we compress the embedding prompt into a pair of low-rank matrices $\mathbf{u} \in \mathbb{R}^{m \times r}$ and $\mathbf{v} \in \mathbb{R}^{n \times r}$, such that the embedding $\mathbf{c} \in \mathbb{R}^{m \times n}$ is reconstructed as:

$$\mathbf{c} = \frac{\mathbf{u} \cdot \mathbf{v}^\top}{\sqrt{r}} \qquad (6)$$

This formulation allows rank $r$ to trade off between bitrate and generation quality, which is set to 8 in our experiment. To reduce quantization loss, quantization is integrated into the gradient descent process, allowing the system to fit directly to the quantized matrices.

For video sequences, addressing the substantial semantic redundancy across consecutive frames is paramount. A naive frame-by-frame inversion is inherently inefficient, as it fails to exploit the intrinsic temporal coherence of video. To overcome this, the encoding framework is augmented to enforce temporal continuity through inter-frame compression. This is achieved by introducing a regularization term to the optimization objective, which penalizes significant deviations between the prompt embeddings of adjacent frames. This methodology ensures that only the semantic innovations between frames are encoded for transmission, dramatically reducing temporal data redundancy and further enhancing overall communication efficiency:

$$\lambda = \|\mathbf{c}_t - \mathbf{c}_{t-1}\|_2 \qquad (7)$$

where $\mathbf{c}_t$ and $\mathbf{c}_{t-1}$ are prompts for consecutive frames. The final loss function becomes:

$$\mathcal{L} = \beta \cdot \mathcal{D} + (1-\beta) \cdot \lambda \qquad (8)$$

with $\beta \in [0,1]$. This regularization ensures that temporally adjacent prompts remain close in the prompt space, enabling intermediate prompts to be approximated by linear interpolation at the receiver side. As a result, only keyframe prompts need to be transmitted, significantly reducing bitrate. In practice, $\beta$ is set to 0.2 and we transmit a prompt every ten frames.

**Calculation of SOP intervals**

The distribution of the densest SOPs on the Poincare sphere is essentially a filling problem for the $S^2$ sphere. In 1978, K. Böröczky [34] investigated the problem of filling the spherical cap on the unit sphere $S^{n-1}$ and proposed an approximate solution:

$$N(n,\theta) \leq \frac{1-\cos\theta}{2\cos\theta} \cdot \left(\frac{4}{1+\sin\theta}\right)^n \qquad (9)$$

where $\theta$ is the arc distance between two adjacent points. $n$ is the dimension, and for a three-dimensional Poincare sphere, $n=3$. Therefore, the above equation can be further simplified:

$$N \sim \frac{\pi}{3}\left(\frac{2}{\theta}\right)^2 \qquad (10)$$

In our experiment, the minimum value of $\theta$ is $\pi/2048$. It is calculated that up to $N = 1.7 \times 10^6$ non-crossed SOPs can be obtained on the sphere.

**Acknowledgement**
This research is supported by the National Key Research and Development Program of China (2023YFB2906300); National Natural Science Foundation of China (62225110 and 62205114); JD project of Hubei province (2023BAA013); Hubei Provincial Natural Science Foundation of China (2025AFB008).

**Data availability**
All data are available from the corresponding author upon request.

**Code availability**
All codes are available from the corresponding author upon request.

**Author contributions**
Y.X., Z.Z., and J.R. contributed equally to this work. M.T. conceived the idea and supervised the research with H.W. and S.Y. Y.X. designed and performed the experiments with J.L. Z.Z., G.Y., Y.L., Z.G., T.J., X. H. and M.Z. provided technical support for the experiments. Z.Z. developed the polarization modulation algorithm and implemented the control code. J.R. designed and implemented the semantic encoding framework. Y.X. and J.R. analyzed the experimental data. Y.X., J.L. and J.R. prepared the figures. Y.X. drafted the manuscript with input from all authors. All authors revised the paper.

**Conflict of interest**
The authors declare no competing interests.